# Scoring the European Citizen in the AI era *(postprint)*

N. Genicot

*Perelman Centre for Legal Philosophy (ULB, Brussels), and Law, Science, Technology & Society (LSTS) Research Group (VUB, Brussels), Belgium*

**Abstract**

Social scoring is one of the AI practices banned by the AI Act. This ban is explicitly inspired by China, which in 2014 announced its intention to set up a large-scale government project – the Social Credit System – aiming to rate every Chinese citizen according to their good behaviour, using digital technologies and AI. But in Europe, individuals are also scored by public and private bodies in a variety of contexts, such as assessing creditworthiness, monitoring employee productivity, detecting social fraud or terrorist risks, and so on. However, the AI Act does not intend to prohibit these types of scoring, as they would qualify as "high-risk AI systems", which are authorised while subject to various requirements. One might therefore think that the ban on social scoring will have no practical effect on the scoring practices already in use in Europe, and that it is merely a vague safeguard in case an authoritarian power is tempted to set up such a system on European territory. Contrary to this view, this article argues that the ban has been drafted in a way that is flexible and therefore likely to make it a useful tool, similar and complementary to Article 22 of the General Data Protection Regulation, to protect individuals against certain forms of disproportionate use of AI-based scoring.

1. **Introduction**

Scores, whether used by public or private actors, are pervasive in our societies. Citizens are scored, often without their knowledge, to determine whether they are creditworthy,[1] good workers,[2] safe drivers,[3] potential terrorists,[4] likely to commit social fraud[5] and so on. Claiming to predict behaviour, these scores are used to make decisions about individuals in key areas. Companies and public authorities either develop scores themselves or buy them from data brokers and credit agencies who collect and analyse vast quantities of data from a variety of sources.

Given that scores are increasingly produced using artificial intelligence ("AI") methods, the question arises as to how they are affected by the recently adopted Artificial Intelligence Act ("AI Act"),[6] and what this regulation changes from existing law. The risk-based approach that structures this new legislation is now well-known: AI systems which present a too substantial risk are banned, those that present a high risk are subject to a series of legal requirements, and those with a low risk are subject to minimal or no requirements. While scoring techniques are not regulated as a whole, one of the few banned uses of AI listed in Article 5 concerns "social scoring".

Compared to other banned AI practices such as real-time biometric identification, social scoring has been the subject of relatively little debate between co-legislators. The only major amendment that has been added to the Commission's initial proposal is to ban social scoring in general, whether implemented by public or private organisations, while the Commission had limited the ban to public authorities.[7] What was clear was that the EU wanted to distinguish itself from China, which in 2014 announced its intention to set up a "Social Credit System", a large-scale government project aiming to rate every Chinese citizen according to their good behaviour, using digital technologies and AI. Although the reality in China is more nuanced – the Social Credit System being far from fully implemented, contrary to what we usually read in the media –,[8] the fact that the EU had this Chinese project in mind when introducing the ban is explicitly acknowledged. The Commissioner for Internal Market Thierry Breton stated in this regard that: "The majority of applications using AI are risk-free. This is not the case for some. Social scoring, for example. China uses it to control individuals. Using data captured without their knowledge, citizens are classified, evaluated and monitored. We are proposing to ban this type of artificial intelligence application in Europe."[9]

---

[1] Mikella Hurley and Julius Adebayo, 'Credit Scoring in the Era of Big Data' [2016] Yale Journal of Law and Technology 148.
[2] Javier Sánchez-Monedero and Lina Dencik, 'The Datafication of the Workplace' (Data Justice Lab, Cardiff University 2019) Working paper; Wolfie Christl, 'Corporate Surveillance in Everyday Life. How Companies Collect, Combine, Analyze, Trade, and Use Personal Data on Billions' (Cracked Labs 2017) <https://crackedlabs.org/en/corporate-surveillance> accessed 14 January 2025.
[3] Wiltrud Weidner, Fabian WG Transchel and Robert Weidner, 'Telematic Driving Profile Classification in Car Insurance Pricing' (2017) 11 Annals of Actuarial Science 213.
[4] Louise Amoore, *The Politics of Possibility: Risk and Security beyond Probability* (Duke University Press 2013).
[5] Vincent Dubois, *Contrôler les assistés. Genèses et usages d'un mot d'ordre* (Raisons d'agir 2021); Lina Dencik and others, 'Data Scores as Governance: Investigating Uses of Citizen Scoring in Public Services' (Data Justice Lab, Cardiff University 2018); Virginia Eubanks, *Automating Inequality: How High-Tech Tools Profile, Police, and Punish the Poor* (First Edition, St Martin's Press 2017).
[6] Regulation (EU) 2024/1689 of the European Parliament and of the Council of 13 June 2024 laying down harmonised rules on artificial intelligence and amending Regulations (EC) No 300/2008, (EU) No 167/2013, (EU) No 168/2013, (EU) 2018/858, (EU) 2018/1139 and (EU) 2019/2144 and Directives 2014/90/EU, (EU) 2016/797 and (EU) 2020/1828 (Artificial Intelligence Act) [2024] OJ L.
[7] This extension was suggested by the European Data Protection Authorities in their opinion on this proposal. The Council of the European Union, in its "general approach" of 25 November 2022, endorsed this view, as did the European Parliament in the text adopted in spring 2023.

However, the AI Act's ban on social scoring is not intended to prevent European citizens from being scored (or "classified, evaluated and monitored", to use Breton's words) in any circumstances. Some forms of AI scoring are not even concerned by the AI Act, such as scoring for advertising purposes.[10] And some other forms of scoring, such as credit scoring, insurance pricing or recidivism risk assessment, qualify as "high-risk AI systems",[11] meaning that they are explicitly authorised while being subject to a number of requirements.

If this ban is not aimed to affect these forms of scoring, one might wonder whether it is likely to have any effect on the uses of AI that are already taking place in the EU, a question that seems all the more relevant that it now concerns both private and public actors. The fact that social scoring is presented as only being used in China suggests that its prohibition is merely seen as a vague safeguard in case an authoritarian power is tempted to set up such a frightening system on EU territory.

Contrary to this view, this Article argues that the boundary between social scoring and high-risk AI scoring systems within the meaning of the AI Act is far from clear-cut. Indeed, Article 5(1)(c), which contains the ban, proves to be drafted broadly, opening the possibility of wide application. In this regard, the ban on social scoring shares common features with the prohibition of automated decisions set out in Article 22 of the General Data Protection Regulation ("GDPR"),[12] and the way the latter has been recently interpreted by the Court of Justice of the European Union ("CJEU") in the recent *SCHUFA Holding AG* case[13] can be taken as a source of inspiration to anticipate the scope of Article 5(1)(c).

In brief, and this assertion will be developed throughout this Article, the "social" nature of a score (which makes it prohibited) is not an inherent characteristic of the score but rather a matter of *context*. One score can be lawful in one case, and qualify as a social score in another because of the way it is used. This has important implications in terms of the rights of data subjects: far from being a useless provision, Article 5(1)(c) of the AI Act can be used as a tool – similar to Article 22 of the GDPR – to protect individuals against some forms of disproportionate uses of AI-based scores.

---

[8] Zeyi Yang, 'China Just Announced a New Social Credit Law. Here's What It Means.' (*MIT Technology Review*, 22 November 2022) <https://www.technologyreview.com/2022/11/22/1063605/china-announced-a-new-social-credit-law-what-does-it-mean/> accessed 14 January 2025; Tong Lam, 'The People's Algorithms: Social Credits and the Rise of China's Big (Br)Other' in Andrea Mennicken and Robert Salais (eds), *The New Politics of Numbers: Utopia, Evidence and Democracy* (Springer International Publishing 2022) 86; Séverine Arsène, 'The Social Credit System in China. Discipline and morals' (2021) 225 Reseaux 55, LXVIII; Daithí Mac Síthigh and Mathias Siems, 'The Chinese Social Credit System: A Model for Other Countries?' (2019) 82 The Modern Law Review 1034.
[9] "Artificial intelligence: in Europe, innovation and safety go hand in hand | Statement by Commissioner Thierry Breton", https://ec.europa.eu/commission/presscorner/detail/en/STATEMENT_23_3344, 18 June 2023.
[10] This assertion would require further analysis, but – in short – advertising and marketing are not included in the list of high-risk AI systems in Annex III.
[11] See Annex III 5(b)(c) and 6(d).
[12] Regulation (EU) 2016/679 of the European Parliament and Council of 27 April 2016 on the protection of natural persons with regard to the processing of personal data and on the free movement of such data, and repealing Directive 95/46/EC (General Data Protection Regulation) [2016] OJ L 119/1.
[13] Judgment of the 7 December 2023, *SCHUFA Holding AG*, Case C-634/21, ECLI:EU:C:2023:957.

To support this claim, the second section of this Article begins by examining how scores were first regulated by data protection law through the prohibition of automated decision-making, a provision that has its roots in the 1978 French "*Informatique et libertés*" law and is now set out in Article 22 of the General Data Protection Regulation ("GDPR").[14] In this regard, the *SCHUFA* case held by the CJEU is discussed. The third section then describes the core principles of the AI Act and how it tackles scoring techniques. Finally, the fourth section focuses on the ban on social scoring by examining the wording of the AI Act as well as the detailed guidelines published by the European Commission on prohibited AI practices.[15] It argues for a broad interpretation that makes this ban a protection against disproportionate uses of AI-based scores.

Drawing on the existing literature on scoring techniques and the risks they pose to fundamental rights,[16] this Article offers the first in-depth discussion of how scoring is addressed by the AI Act and, in particular, of the implications of the ban on social scoring in light of existing scoring practices.

## 2. Scores and automated decisions in the GDPR

Scoring systems aim to assess or predict a trait relating to an individual (or a group of individuals) – such as creditworthiness or health status – using data analytics and AI.[17]

Schematically, a scoring system is developed in the following way: a database is built up, for example of previous borrowers if the aim is to assess creditworthiness. This data may include information such as credit history, whether they have a bank account, type of employment, professional experience, marital status, etc. The data is then divided in two parts – those who have duly repaid their loans and those who have not – in order to identify the combination of characteristics that tend to predict the target value – in this case, creditworthiness. Each variable is therefore weighted according to whether it correlates with being a good or bad borrower. In the end, the scoring system enables a new applicant to be rated on the basis of his or her own characteristics. The higher the score, the more likely the applicant is to repay the loan.

---

[14] Regulation (EU) 2016/679 of the European Parliament and Council of 27 April 2016 on the protection of natural persons with regard to the processing of personal data and on the free movement of such data, and repealing Directive 95/46/EC (General Data Protection Regulation) [2016] OJ L 119/1.

[15] European Commission, 'Annex to the Communication to the Commission Approval of the content of the draft Communication from the Commission - Commission Guidelines on prohibited artificial intelligence practices established by Regulation (EU) 2024/1689 (AI Act)', C(2025) 884 final ('Guidelines on prohibited AI practices'). The publication of these guidelines was required by Article 96(1)(b) of the AI Act. Note that a draft version of this article – which was submitted in September 2024 – was shared in August 2024 with the team at the AI Office (European Commission) working on these guidelines, and that some statements from this article have been directly reproduced in the guidelines. Where this is the case, it is indicated in footnotes.

[16] See *inter alia*: Francesca Palmiotto, 'When Is a Decision Automated? A Taxonomy for a Fundamental Rights Analysis' (2024) 25 German Law Journal 210; Dencik and others (n 5); Tal Zarsky, 'Understanding Discrimination in the Scored Society' (2014) 89 Washington Law Review 1375; Danielle K Citron and Frank Pasquale, 'The Scored Society: Due Process for Automated Predictions' (2014) 89 Washington Law Review 1; Mireille Hildebrandt and Serge Gutwirth (eds), *Profiling the European Citizen: Cross-Disciplinary Perspectives* (2008). The title of this article deliberately echoes that of the classic book *Profiling the European Citizen* just mentioned.

[17] See Dencik and others (n 5) 10. Legal entities are also scored by commercial credit agencies and credit rating agencies (as regards financial markets). However, since both the GDPR and the ban on social scoring concern decisions made about natural persons, we leave aside legal entities.

Although scoring systems largely predate the emergence of contemporary AI techniques,[18] the rise of Big Data and machine learning has given them considerable importance in a wide range of fields and stimulated their mass adoption by private and public actors.[19] By contrast with traditional statistical techniques, AI models can be fed by a large number of variables that often come directly from people's behaviour as recorded by computers, mobile phones and ubiquitous connected devices. Organisations such as credit institutions, telecom firms, insurance companies, retailers and major platforms score individuals for a variety of purposes, including targeted advertising, credit and risk assessment, employee productivity monitoring, and so on.[20] Scoring systems have also attracted public authorities at national or local level which use them to detect phenomena such as risk of child abuse, welfare fraud, recidivism, terrorism, irregular immigration, etc.[21]

Scores and scoring systems as such are not defined in EU law. However, EU data protection law contains the notion of profiling which closely aligns with scoring. The GDPR defines it as "any form of automated processing of personal data consisting of the use of personal data to evaluate certain personal aspects relating to a natural person, in particular to analyse or predict aspects concerning that natural person's performance at work, economic situation, health, personal preferences, interests, reliability, behaviour, location or movements".[22] The main provision of the GDPR dealing with profiling is Article 22 which prohibits automated decisions. It states that: "the data subject shall have the right not to be subject to a decision based solely on automated processing, *including profiling*, which produces legal effects concerning him or her or similarly significantly affects him or her". The second paragraph of the same provision, though, provides for important exceptions to this prohibition, automated decisions being lawful if they are necessary for the conclusion or performance of a contract, authorised by law or based on the explicit consent of the data subject. However, even when these exceptions apply, Article 22(2)(b) and (3) require the data controller to implement "suitable measures to safeguard the data subject's rights and freedoms and legitimate interests" and "at least the right to obtain human intervention on the part of the controller, to express his or her point of view and to contest the decision". This involves the right to get "meaningful information about the logic involved [in the automated decision], as well as the significance and the envisaged consequences of such processing for the data subject".[23]

---

[18] See *inter alia* in the banking, criminal justice and insurance sectors: Josh Lauer, *Creditworthy: A History of Consumer Surveillance and Financial Identity in America* (Columbia University Press 2017); Daniel Bouk, *How Our Days Became Numbered: Risk and the Rise of the Statistical Individual* (University of Chicago Press 2015); Bernard E Harcourt, *Against Prediction: Profiling, Policing, and Punishing in an Actuarial Age* (University of Chicago Press 2007).
[19] There literature on the contemporary importance of scoring techniques is now very large. See notably: Marion Fourcade and Kieran Joseph Healy, *The Ordinal Society* (Harvard University Press 2024); Dencik and others (n 5); Eubanks (n 5); Citron and Pasquale (n 16); Marion Fourcade and Kieran Healy, 'Classification Situations: Life-Chances in the Neoliberal Era' (2013) 38 Accounting, Organizations and Society 559.
[20] Antonio Aloisi and Valerio De Stefano, *Your Boss Is an Algorithm: Artificial Intelligence, Platform Work and Labour* (Hart 2022); Ifeoma Ajunwa, 'Algorithms at Work: Productivity Monitoring Applications and Wearable Technology as the New Data-Centric Research Agenda for Employment and Labor Law' (2018) 63 Saint Louis University Law Journal 21; Christl (n 2).
[21] See Julia Angwin and others, 'Machine Bias' (*ProPublica*, 23 May 2016) <https://www.propublica.org/article/machine-bias-risk-assessments-in-criminal-sentencing> accessed 16 August 2023; Eubanks (n 5); Dencik and others (n 5); Sam Desiere, Kristine Langenbucher and Ludo Struyven, 'Statistical Profiling in Public Employment Services: An International Comparison', vol 224 (2019) OECD Social, Employment and Migration Working Papers 224; Dubois (n 5); Charly Derave, Nathan Genicot and Nina Hetmanska, 'The Risks of Trustworthy Artificial Intelligence: The Case of the European Travel Information and Authorisation System' (2022) 13 European Journal of Risk Regulation 389. The notion of "risk assessment tool" is largely synonymous with scoring system, but it emphasizes that what is being assessed is framed as a risk (e.g., the risk of offending or irregular immigration). In contrast, scoring systems have a broader scope, as they can be used to predict outcomes that are not necessarily undesirable (e.g., in marketing, predicting who is likely to click on a certain link).
[22] Article 4(4) of the GDPR.

Automated decisions about individuals, whether by private actors or public authorities, often involve profiling and scoring, as they typically assess the likelihood of a person having a certain characteristic (e.g., being creditworthy, a welfare fraudster, or a recidivist). Scoring systems calculate this probability, and the decision is made by setting a threshold: if the score exceeds this threshold, the individual will be considered to possess this quality (e.g. creditworthy) and will be treated accordingly (e.g. receive a credit).

A detour through the history of this provision confirms this strong link between scores and automated decisions. Article 22 of the GDPR is the successor to Article 15 of Directive 95/46/EC,[24] which was itself directly inspired by the French "*Informatique et libertés*" law of 6 January 1978, one of the first data protection laws adopted in Europe. Article 2(2) of this law stipulated that "[n]o administrative or private decision involving an assessment of human behaviour may be based solely on automated processing of information giving a definition of the profile or personality of the person concerned".[25]

The *raison d'être* for this provision in the French law was to be found in projects run by the public authorities, in particular the *GAMIN* system (for "*gestion automatisée de médecine infantile*"), a scoring system designed in 1970 by the Ministry of Health to prevent disabilities in children.[26] Using data from health certificates issued by doctors during medical check-ups, it was designed to automatically identify children at risk who required medical or social assistance. The information included data such as the parents' occupation. When the scoring system (which was not based on AI but on more traditional statistical methods) identified a child as being at risk, this could lead, for example, to compulsory referral to special education programmes. However, the most precarious members of the population were more likely to see their children receive this label, which provoked strong criticism and played a decisive role in the introduction of a ban on solely automated decisions.[27] In the 1980s, the discussions surrounding Article 2 of the French law also related to scoring – in particular the credit scores produced by lending institutions which were closely scrutinised by the *Commission nationale de l'Informatique et des libertés* ("CNIL"). In its 1985 report, the CNIL noted that "decisions to grant or refuse credit to finance the purchase of goods or services are taken on the basis of [a] score" and questioned the legality of this practice with regard to the prohibition of automated decisions.[28]

---

[23] Articles 13(2)(f), 14(2)(g) and 15(1)(h) of the GDPR.
[24] Directive 95/46/EC of the European Parliament and of the Council of 24 October 1995 on the protection of individuals with regard to the processing of personal data and on the free movement of such data (O.J. L 281, 23.11.1995, p. 31).
[25] Our translation.
[26] Colette Hoffsaes, 'Le Système GAMIN. Erreur Technocratique Ou Premier Pas Vers Un Fichage Généralisé ?' [1982] Esprit 22.
[27] Albert Salgueiro, 'Les Modes d'évaluation de La Dignité de Crédit d'un Emprunteur' (Université d'Auvergne - Clermont-Ferrand I 2004) 535. 'Gamin' means kid in French.

Following the migration of this provision (with some amendments) into Directive 95/46/EC and then into the GDPR, and with the rise of algorithmic and AI systems, the prohibition of automated decisions has become all the more topical. The provision contains two main criteria which both have given rise to considerable doctrinal debate: i) the decision about the data subject must produce legal or significant effects; ii) the decision must be solely automated.

First, regarding the *legal or significant effects*, Recital 71 of the GDPR illustrates these effects by mentioning the refusal of a credit application and online recruitment practices. The Article 29 Working Party guidelines written on profiling and automated decision-making – which have only doctrinal value – provides more examples.[29] In terms of legal effects, the guidelines refer to events such as the cancellation of a contract, the refusal of a social benefit granted by law, or the refusal of a residence permit in a country. In terms of significant effects, the guidelines specify that this should refer to effects which "significantly affect the circumstances, behaviour or choices of the individuals concerned; have a prolonged or permanent impact on the data subject; or at its most extreme, lead to the exclusion or discrimination of individuals".[30] Examples include access to health services or to educational institutions.

Second, when is a decision *solely* automated?[31] Although it may seem obvious at first glance, it is difficult to define a clear criterion for distinguishing a decision that has been automated from one in which a human being is involved. Indeed, decision-making systems concerning individuals are rarely fully automated. Human agents are often involved at one stage or another, but this doesn't mean that algorithms can't play a decisive role. The Article 29 Working Party considers that the condition of human involvement implies that the oversight of the decision must be "meaningful, rather than just a token gesture".[32] But what if, for example, an automated decision-making system identifies people suspected of having committed welfare fraud, which then triggers a control by government officials. The control itself would be carried out by a human being, but it would have been triggered because of a prior automated decision. And what if, to take another example, a human agent takes a final individual decision but relies heavily on the output of an algorithm? How does one determine whether the agent's involvement is sufficiently meaningful?

---

[28] Commission nationale de l'informatique et des libertés, '6ᵉ Rapport d'activité. 1er Janvier 1985 -' (La documentation française 1986) 82. Our translation.
[29] Article 29 Working Party, 'Guidelines on Automated Individual Decision-Making and Profiling for the Purposes of Regulation 2016/679' (2018) <https://ec.europa.eu/newsroom/article29/items/612053> accessed 12 July 2024.
[30] ibid 21.
[31] This criterion has perhaps generated even more discussion. Article 29 Working Party has written guidelines on profiling and automated-decision making (which had been amended by the European Data Protection Board): Article 29 Working Party (n 29). See also : Palmiotto (n 16); Reuben Binns and Michael Veale, 'Is That Your Final Decision? Multi-Stage Profiling, Selective Effects, and Article 22 of the GDPR' (2021) 11 International Data Privacy Law 319; Lee A Bygrave, 'Article 22 Automated Individual Decision-Making, Including Profiling' in Christopher Kuner and others (eds), *The EU General Data Protection Regulation (GDPR): A Commentary* (Oxford University Press 2020); Lee A Bygrave, 'Automated Profiling: Minding the Machine: Article 15 of the EC Data Protection Directive and Automated Profiling' (2001) 17 Computer Law & Security Review 17.
[32] Article 29 Working Party (n 29) 21.

Importantly, the first major case in which the CJEU had the opportunity to rule on the scope of this provision is the case C-634/21 *SCHUFA Holding AG* which only dates back to 2023.[33] It precisely relates to scores, namely a credit score produced by the German consumer credit agency SCHUFA. This company was founded in 1927, when the German consumer credit market was being nationalised. In the 1970s, it had almost 22 million files, representing around 80% of the country's working population and it remains today the main credit bureau in the country.[34] The dispute arose from a SCHUFA credit score sold to a third party (a bank or other type of credit institution) which led this third party to refuse a loan to the applicant. After being refused the loan, the applicant asked SCHUFA to disclose information about the data stored and to erase incorrect data. In response, SCHUFA informed her that her score was 85.96%, and provided some general information about the methodology used, but refused to say more, claiming business confidentiality. The applicant then sued SCHUFA, relying on various provisions of the GDPR, in particular Article 22.

The main issue was indeed the applicability of article 22 as SCHUFA itself does not grant any loans nor provide any services to individuals, it merely offers information to commercial actors. Therefore, strictly speaking, SCHUFA does not take any decisions concerning anyone as it has no contractual relationship with consumers. However, depending on how credit institutions interpret a score, its influence varies. A score may be only one of many factors considered or, conversely, it may be the only criterion taken into account by the lending institution. In many cases, a lending institution will not base its decision solely on the credit bureau score but will automatically refuse to grant credit to applicants with a score below a certain threshold. In such a case, it is difficult to determine where exactly the decision is taken. Is the decision taken solely by the credit institution, or is the SCHUFA score itself a decision?

To answer this question, the CJEU relies on the description given by the German referring court, according to which the third party draws "strongly" on credit scores so that an insufficient SCHUFA score leads "in almost all cases, to the refusal of that bank to grant the loan applied for".[35] This factual finding demonstrates, according to the Court, that the score itself constitute a decision even though the entity producing the credit score is not the lending institution.

Scores are therefore not explicitly mentioned in the GDPR, but they are nonetheless directly affected by the ban on automated decisions. A score which strongly leads to certain consequences corresponding to "legal or significant effects" is prohibited, unless exceptions apply (such as an explicit law authorising this automated decision). However, as the CJEU found in the SCHUFA case, the "decision-making" nature of a score is a matter of context and is not defined *ab initio* when the score is designed. The same score may be classified as an automated decision in certain circumstances, but not in others.

---

[33] Judgment of the 7 December 2023, *SCHUFA Holding AG,* Case C-634/21, ECLI:EU:C:2023:957. For a brief commentary: Nathan Genicot, 'To Score Is to Decide. About the SCHUFA Case' [2023] Verfassungsblog <https://verfassungsblog.de/to-score-is-to-decide/> accessed 1 July 2024.
[34] Larry Frohman, 'Virtually Creditworthy: Privacy, the Right to Information, and Consumer Credit Reporting in West Germany, 1950–1985' in Jan Logemann (ed), *The Development of Consumer Credit in Global Perspective* (Palgrave Macmillan US 2012).
[35] Para 48.

In addition to EU data protection law, scoring techniques will now have to comply with the AI Act. The question we must now address, therefore, is how this new legislation deals with scoring techniques and to what extent it complements the prohibition on automated decisions already present in the GDPR. This is examined in the following two sections.

### 3. Scores in the AI Act

To understand how the AI Act addresses scoring techniques, it is useful to take a look at the *Ethics Guidelines for Trustworthy AI* ("*Ethics Guidelines*"), published in 2019 by the High-Level Expert Group on Artificial Intelligence set up by the European Commission and which had a strong influence on the Commission's regulation proposal.[36]

Scoring techniques are discussed in the document among other examples of AI use that raise critical concerns, and are described as endangering the principles of autonomy and non-discrimination. However, the *Ethics Guidelines* draw a distinction between "citizen scoring – on a large or smaller scale – [that] is already often used in purely descriptive and domain-specific scoring" and "normative citizen scoring (general assessment of "moral personality" or "ethical integrity") in *all* aspects and on a large scale" which presents a risk for democratic values, especially "when used disproportionately and without a delineated and communicated legitimate purpose".[37]

What seems to distinguish this normative citizen scoring from other types of scoring is 1) that it consists of a general assessment of the individual's personality and 2) that this assessment is not specific to one area or purpose but has consequences for all aspects of life, and is used disproportionately. The first point concerns what might be called the stage of "score design", i.e. the way in which the score is developed and what it is intended to assess. And the second point concerns what might be called the sage of "score use", i.e. the uses made of the score and the consequences attached to it.

To some extent, this distinction between normative citizen scoring and other types of scoring correspond to the distinction that can be found in the AI Act between social scoring and high-risk AI scoring systems. As we know, this regulation is both general in its application, since in principle it covers all types of AI systems, and risk-based, since the level of requirements and obligations imposed on the AI system depends on the risk the system poses to health, safety and fundamental rights. While certain so-called "AI practices" are prohibited because the risk they pose is deemed "unacceptable", other AI systems are classified as "high-risk" and therefore authorised but subject to a series of requirements.[38] Certain other AI systems, such as chatbots, which present a low risk, are subject to transparency and information requirements.[39]

---

[36] High-Level Expert Group on Artificial Intelligence, 'Ethics Guidelines for Trustworthy AI' (European Commission 2019).
[37] ibid 34. As explained in Introduction, this normative citizen scoring clearly refers to the Chinese social credit system project.

In line with this logic, AI-based scores are regulated differently depending on their level of risk. First, "social scoring" is one of the prohibited practices set out in Article 5(1)(c) of the AI Act. Reproducing the distinction between the design of the score (first paragraph) and the use of the score (points (i) and (ii) of the second paragraph), this provision states that:

> Are prohibited "the placing on the market, the putting into service or the use of AI systems for the purpose of the evaluation or classification of natural persons or groups of persons over a certain period of time based on their social behaviour or known, inferred or predicted personal or personality characteristics, with the social score leading to either or both of the following:
>
> (i) detrimental or unfavourable treatment of certain natural persons or whole groups of persons in social contexts that are unrelated to the contexts in which the data was originally generated or collected;
>
> (ii) detrimental or unfavourable treatment of certain natural persons or groups of persons that is unjustified or disproportionate to their social behaviour or its gravity."[40]

In addition, there are also traces of scoring systems in the list of high-risk AI systems. While the word "score" is only mentioned once to refer to credit scoring, a closer look at Annex III leads to the conclusion that many other high-risk AI systems in fact consist of a scoring system. Without being exhaustive, we can mention AI systems used to recruit new employees or assess their performance once they have been hired,[41] to select and rate insurers for life and health insurance,[42] to assess the eligibility of citizens for public assistance benefits,[43] to assess the risk of recidivism[44] or to assess the risk of illegal immigration.[45] As explained in the previous section, all these applications are likely to rely on scoring methods.[46]

---

[38] The classification rules for high-risk AI systems are contained in Article 6 which refers to Annexes of the Regulation. One the one hand, a list of different types of AI systems lies in Annex III. On the other hand, AI systems that are used as a safety component of a product (or are themselves a product) covered by the Union harmonisation legislation listed in Annex I, and that require third-party conformity assessment pursuant to this legislation are also classified as high-risk. There are also specific rules for general-purpose AI models (See Chapter V of the AI Act).
[39] Most of the rules contained in the AI Act will apply from 2 August 2026, but the rules relating to prohibited practices came into force on 2 February 2025.
[40] In addition, Recital 31 precises that "AI systems providing social scoring of natural persons by public or private actors may lead to discriminatory outcomes and the exclusion of certain groups. (…) Such AI systems evaluate or classify natural persons or groups thereof on the basis of multiple data points related to their social behaviour in multiple contexts or known, inferred or predicted personal or personality characteristics over certain periods of time. (…) AI systems entailing such unacceptable scoring practices and leading to such detrimental or unfavourable outcomes should be therefore prohibited. That prohibition should not affect lawful evaluation practices of natural persons that are carried out for a specific purpose in accordance with Union and national law."
[41] Annex III (4)(a) of the AI Act.
[42] Annex III (5)(c) of the AI Act.
[43] Annex III (5)(a) of the AI Act.

On the one hand, the legal regime applicable to the ban on social scoring is simple: if a score falls within its scope, it is unlawful. The underlying logic is therefore very similar to that of Article 22 of the GDPR: like automated decisions, social scores are prohibited.[47] In contrast, the rules imposed on high-risk AI systems are very complex. These include establishing a risk management system, drafting technical documentation which demonstrates compliance with the AI Act, maintaining a data governance framework (which notably aims to control the quality and representativeness of the data used to feed the AI model), ensuring that a human oversees the AI system and its outputs, etc.[48] Before placing a "high-risk" AI system on the market or putting it into service, the providers of such systems will have to carry out a conformity assessment to ensure that they meet all these obligations.

The ban on social scoring is thus not intended to affect licit uses of scores, including high-risk AI scoring systems. In this sense, the last sentence of Recital 31 of the AI Act states that the "prohibition [of social scoring] should not affect lawful evaluation practices of natural persons that are carried out for a specific purpose in accordance with Union and national law".[49] However, the boundary between social scoring and other forms of scoring is not necessarily straightforward. As underlined by the European Commission in its Guidelines on prohibited AI practices, "the use of AI systems classified as high-risk may in some cases qualify as a prohibited practice in specific instances if all conditions under one or more of the prohibitions in Article 5 AI Act are fulfilled".[50] To understand when a score risks being prohibited under Article 5(1)(c), it is therefore essential to take a close look at the conditions set out in this provision.

### 4. How to distinguish social scoring from lawful scoring?

Using the distinction between score design and score use, this final section analyses Article 5(1)(c) and compares it to existing scoring practices. This will enable us to test the robustness of the distinction between social scoring within the meaning of the AI Act and other forms of scoring and to determine what the foreseeable effects of the ban will be. By following a step-by-step analysis, the analysis will show that the scope of the ban is broad and likely to cover different types of scoring devices.

---

[44] Annex III (6)(d) of the AI Act.
[45] Annex III (7)(b) of the AI Act.
[46] It should be noted that AI systems listed in Annex III will not automatically be classified as high-risk given that a provision, Article 6(3), was added during the legislative process allowing providers to escape from this qualification in some circumstances (for instance when the AI system only performs a narrow procedural task). But interestingly, the last paragraph adds that AI systems listed in Annex III "shall always be considered to be high-risk where the AI system performs profiling of natural persons". This precision is important since scoring is always a form of profiling, as explained in the previous section. This means that a scoring system falling under one of the AI systems listed in Annex III should always be considered as high risk.
[47] However, as discussed in the conclusion, unlike Article 22 of the GDPR, Article 5(1)(c) of the AI Act does not provide for any exceptions.
[48] See Chapter 2 of the AI Act.
[49] See also Guidelines on prohibited AI practices (n 15) 61, para. 175.
[50] Guidelines on prohibited AI practices (n 15) 12, para. 37.

## 4.1. The stage of score design

Simply put, the first paragraph of Articles 5(1)(c) defines "social score" as the evaluation or classification of natural persons based on their "social behaviour" or "known, inferred or predicted personal or personality characteristics".

This definition is close to the definition of profiling set out in EU data protection law which was discussed in Section 2.[51] Both definitions refer to the evaluation of a natural person.[52] One major relevant difference is that the AI Act only concerns AI systems while the GDPR mentions automated processing of data, which is not necessarily based on AI.

Another important difference is that the definition of profiling requires the evaluation to be based on personal data, while the definition of social scoring makes no reference to personal data but rather to social behaviour, personality characteristics and personal characteristic. None of these concepts are defined in the regulation.[53]

What is behaviour and, subsequently, what is "social behaviour" (as opposed to behaviour that is not social)? If we think of digital traces, should we consider that interactions on social media are social behaviour, but that browsing history is not? The entire history of the social and behavioural sciences has been marked by discussions on the meaning and limits of these two concepts, and the least we can say is that there is no clear, unequivocal understanding of them.

As for the concept of "personality characteristic", it has its origin in psychology, which since the 19th century has attempted to categorise and classify human beings on the basis of their mental aptitudes or personality. From the outset, psychometric tests were designed to distinguish between children on the basis of their aptitudes or to select and guide job applicants.[54] Today, there is a great deal of work looking at the use of personality traits for predictive purposes. Some studies examine the psychological determinants of employability (which is considered useful for public services when dealing with the unemployed).[55] In the field of credit too, psychometrics is often used. The company LenddoEFL – which specialises in the use of alternative data for credit assessment – explains on its website that "a psychometric credit assessment provides an alternative for thin-file loan applicants (*i.e.,* zero or low credit history) by generating credit scores based on personality and behavior".[56] Other research attempts to show how various sources of data can be used to predict personality traits. In a 2013 article, researchers studied the predictive power of Facebook likes with regard to various personal characteristics, including the Big Five personality dimensions, intelligence and even sexual orientation.[57] This is all the more relevant given that Article 5(1)(c) specifies that not only "known" but also "inferred or predicted" personality characteristics are concerned.

---

[51] In the same sense, see Guidelines on prohibited AI practices (n 15) 52, para. 154.
[52] As the European Commission's Guidelines on prohibited AI practices point out, the notion of classification (which is also included in the definition) is even broader than the notion of evaluation, since it includes assignment to a category such as age or gender (Guidelines on prohibited AI practices (n 15) 52, para 153). See also Article 29 Working Party (n 29) 7. It also worth noting that scores should be distinguished from individual ratings by which consumers assess the quality of a service (such as an Uber driver, or a restaurant) as these ratings do not involve AI but consist of a mere aggregation of individual preferences. AI-enabled scoring on its part is not a matter of summing up individual preferences, but of assessing the probability of future behaviour. The European Commission's guidelines align with this perspective (see Ibid. 61).
[53] The term "personality characteristic" is also mentioned in the text in relation to AI systems intended to be used by law enforcement authorities or on their behalf to assess personality traits and characteristics of natural persons or groups (See Article 5(1)(d) and Annex III, (6)(d)).
[54] For an overview of the work carried out in this field at the time: JM Lahy, 'Les Profils Psychologiques Dans La Sélection et l'orientation Professionnelle' [1926] Prophylaxie mentale 178; Hugo Münsterberg, *Psychology and Industrial Efficiency* (The Riverside Press 1913).
[55] Claude Houssemand, Anne Pignault and Raymond Meyers, 'A Psychological Typology of Newly Unemployed People for Profiling and Counselling' (2014) 33 Current Psychology 301; Selver Derya Uysal and Winfried Pohlmeier, 'Unemployment Duration and Personality' (2011) 32 Journal of Economic Psychology 980; Majella J Albion, Karen M Fernie and Lorelle J Burton, 'Individual Differences in Age and Self-Efficacy in the Unemployed' (2005) 57 Australian Journal of Psychology 11. Uysal and Pohlmeier show, for example, that of the different personalities in the famous Big Five model, the 'conscientious mind' personality type has a high probability of finding a job, while the 'neuroticism' personality type has little chance of finding a job quickly.

Lastly, "personal characteristic" is probably the broadest of the three notions.[58] It seems very close to the concept of personal data that is defined in Article 4(1) of the GDPR as "any information relating to an identified or identifiable natural person ('data subject')" and has been interpreted in a flexible and extensive way by the CJEU.[59] This broad definition and the ongoing process of datafication of all aspects of society have led some to conclude that almost any data can be personal data, making data protection law the "law of everything".[60] The notion of personal characteristic could be considered to have an even fuzzier meaning than that of personal data, since it is not defined in the text and the condition of "identifiability" contained in the definition of personal data is therefore not required. In this sense, it would be sufficient for a piece of information to relate to an individual for it to constitute a personal characteristic.

Given all this, it does not seem excessive to conclude that all types of data are covered by Article 5(1)(c). As a result, it is really difficult to imagine an individual AI-based score which would not qualify as a "social score" under this definition. One might even argue that every time an AI system evaluates or classifies someone, it performs "social scoring" within the meaning of the AI Act. This definition however is problematic because it corresponds to any form of scoring and it is hard to see what exactly makes the scoring *social*. Think of credit scoring already mentioned: it is the evaluation of someone on the basis of certain personal characteristics (such as income, age, family situation, consumer habits, etc.). This is precisely the definition of social scoring given in the AI Act.

The definition of "social score" was presumably intended to refer to AI systems assessing a person's "morality", "integrity" or "trustworthiness". But in the end, looking at the precise wording of the provision, it appears to encompass any form of individual scoring. As far as the score design stage is concerned, there is therefore nothing that distinguishes social scoring from lawful scoring. What remains to be examined is the question of the uses to which a score may be put.

---

[56] 'LenddoEFL' (*LenddoEFL*, 17 May 2021) <https://lenddoefl.com> accessed 16 July 2024. In a 2013 book, economists from this company describe the value of psychometric tests for credit development (Bailey Klinger, Asim Ijaz Khwaja and Carlos del Carpio, *Enterprising Psychometrics and Poverty Reduction* (Springer New York 2013)).
[57] Michal Kosinski, David Stillwell and Thore Graepel, 'Private Traits and Attributes Are Predictable from Digital Records of Human Behavior' (2013) 110 Proceedings of the National Academy of Sciences 5802.
[58] The European Commission's Guidelines on prohibited AI practices state that personality characteristics and personal characteristics should be considered synonymous, but there is no reason why this should be the case. In our view, personal characteristics have a broader scope than personality traits. See Guidelines on prohibited AI practices (n 15) 54, para. 159.
[59] Lee A Bygrave and Luca Tosoni, 'Article 4(1). Personal Data' in Christopher Kuner and others (eds), *The EU General Data Protection Regulation (GDPR): A Commentary* (Oxford University Press 2020).
[60] Nadezhda Purtova, 'The Law of Everything. Broad Concept of Personal Data and Future of EU Data Protection Law' (2018) 10 Law, Innovation and Technology 40.

### 4.2. The stage of score use

The second part of Article 5(1)(c) provides that social scores are banned only a) if they lead to "detrimental or unfavourable treatment" b) in "social contexts that are unrelated to the contexts in which the data was originally generated or collected" and/or c) "that is unjustified or disproportionate to their social behaviour or its gravity".

#### 4.2.1. A score leading to a detrimental treatment

This criterion conceals two questions: what is a detrimental or unfavourable treatment? And when does a score *lead* to a detrimental or prejudicial treatment? These interrogations are strikingly similar to those raised by Article 22 of the GDPR.

First, the condition of a detrimental of unfavourable treatment echoes the "legal or significant effects" of Article 22 discussed in Section 2 since the expression "detrimental of prejudicial treatment" used in the AI Act, although vague, also refers to the impact of the score on the situation of the scored subject, which corresponds more broadly to the logic behind the AI Act's risk-based approach, i.e. the regulation of AI systems according to their effects. Recital 31 of the AI Act also states that social score may lead to "discriminatory outcomes and the exclusion of certain groups" and "violate the right to dignity and non-discrimination and the values of equality and justice". Violation of the right to dignity or non-discrimination is therefore not a condition in itself for finding the existence of a prohibited social score within the meaning of the AI Act, but it does constitute the *raison d'être* underpinning this prohibition. In this regard, it is noteworthy that the examples of "legal or significant effects" given by the Article 29 Working Party are very similar to the types of AI use listed as high-risk in the AI Act (i.e. AI used in the workplace, in educational institutions, in the banking sector, in social welfare, etc.).[61]

Second, Article 5(1)(c) of the AI Act provides that social scoring must *lead* to a detrimental or unfavourable treatment in order to be prohibited. In other words, the treatment must be the consequence of the score, and the score the cause of the treatment.[62] This parallels the "sole automation" criterion of Article 22 of the GDPR. Here again, the wording is less precise than Article 22 of the GDPR since it is not specified that the social score must "solely" lead to the treatment, which gives room for a broader interpretation. But we can assume that if an AI score is only one element among many others leading to the treatment, the provision is less likely to apply. The score should play a decisive role in order to fall under the scope of Article 5(1)(c). However, the conclusion of the *SCHUFA Holding CA Case*, namely that a score produced by an institution other than the one taking the final decision can constitute an automated decision, should be transposed: a score could "lead to a detrimental or unfavourable treatment" even if this score was not produced by the organisation which is the author of the treatment, but by another organisation.

---

[61] One difference with Article 22 is that the treatment must be detrimental or unfavourable, while Article 22 does not explicitly require the decision to be negative but only to produce legal or significant effects.
[62] This analysis has been endorsed by the European Commission in its Guidelines on prohibited AI practices. See Guidelines on prohibited AI practices (n 15) 55, para. 160-162.

*4.2.2. The variety of application contexts*

Finally, one of two alternative criteria (points i) or ii) of the provision) is required for a score to fall within the scope of the ban. The first of these two alternative criteria concerns the context of application: the detrimental treatment must have occurred in social contexts unrelated to the contexts in which the data was originally collected or generated. This condition implicitly intends to prohibit the use of a score for multiple purposes. As stated in the *Ethics guidelines*, social scoring is supposed to concern "all aspects". In the same vein, Recital 17 of the Commission's initial AI Act proposal assumes that social scoring is done "for general purpose". This has been removed in the final version of the AI Act but the last sentence of the actual Recital 31 specifies that the ban does not affect lawful scoring carried out for a "specific purpose", implying *a contrario* that scores that are not used for a specific purpose should be prohibited.[63]

However, the condition that the context of data collection is unrelated to the context in which the score is used is not identical to the condition that the social score is carried out for general-purpose. The former is arguably broader than the latter.

Considering a score designed and used for a specific purpose such as credit scoring, what if the data used to build a scoring system appeared very loosely related to creditworthiness? Could we consider the data feeding the score to be unrelated to its context of application (i.e. granting credit). Classic scores such SCHUFA scores are fed by information such as payment defaults, claims, credit history, etc. which are all more or less related to credit. But new forms of credit scoring include various source of information such as the time at which the credit application is submitted online or activity on social networks, a tendency well illustrated by the expression "all data is credit data".[64] In this case, could it not be considered that the data fed into the score is "unrelated to its context of application" (i.e. the granting of credit), even though the score is used for a specific purpose?[65]

In addition, scores are sometimes used for purposes beyond those for which they were originally designed. This is particularly true of credit scores produced by agencies such as Equifax, Experian, and TransUnion in countries like the United States, Canada, and the United Kingdom. Initially, these scores were intended for banks assessing loan applicants. However, they soon came to be consulted by landlords selecting tenants, employers hiring staff, and car insurance companies determining premiums.[66] In Canada, credit scoring has been characterised as a "parallel justice system", functioning as "a default judgment, which is based solely on the creditors' version of the facts" and does not respect the general legal principle *audi alteram partem*.[67] Once produced, a score can thus be used for purposes that were not initially intended by those who designed them, a phenomenon that sociologist Akos Rona-Tas calls the "off-label" use of scores.[68]

---

[63] This condition echoes the purpose limitation principle set out in Article 5(1)(b) of the GDPR which requires that data must be "collected for specified, explicit and legitimate purposes and not further processed in a manner that is incompatible with those purposes". However, we will not proceed to an exhaustive comparison of these two provisions in this Article.
[64] Rob Aitken, '"All Data Is Credit Data": Constituting the Unbanked' (2017) 21 Competition & Change 274.
[65] In its Guidelines on prohibited AI practices, the European Commission gives the example of a tax authority that would take into account taxpayers' Internet connections to decide who to inspect, and considers that such practices would fall within the scope of the prohibition (Guidelines on prohibited AI practices (n 15) 57).
[66] Akos Rona-Tas, 'The Off-Label Use of Consumer Credit Ratings' (2017) 42 Historical Social Research / Historische Sozialforschung 52.

Credit scores used in this way contribute to determining the destiny of individuals in different contexts which are crucial to everyday life. More than just predictors of an individual's financial stability, they are used as true indicators of their personality, seriousness and integrity. In his history of creditworthiness, historian Josh Lauer emphasises this point. He explains that the overlap between traits such as creditworthiness, health and employability reflected "an implicit understanding that creditworthiness – one's trustworthiness as a financial subject – was conceptually linked to moral qualities involving one's lifestyle, reputation, physical fitness, and social conformity".[69] From a legal perspective, should such use of credit scores for purposes other than granting loan take place in an EU Member State, it could be prohibited under Article 5(1)(c)(i) of the AI Act despite the fact that scoring system was not intended to be "social". The important point is thus that the intention that existed when producing the score is irrelevant, only the practical consequences of the score matter. If a score happens to be bought and used by a company or public service in a context other than the one in which it was originally produced, it could qualify as a social score.[70]

*4.2.3. The disproportionate punishment of a social behaviour*

The second alternative criterion relates to proportionality: a social score shall also be prohibited if it leads to a detrimental or unfavourable treatment that is unjustified or disproportionate to the social behaviour of the scored subject or its gravity. As already seen, the first paragraph of Article 5(1)(c) specifies that social scores are based on social behaviour, personality or personal characteristics, which are all very broad notions, encompassing almost all kinds of data. Here however, the focus is solely on social behaviour. Why is that? What seems to differentiate "social behaviour" from personality of personal characteristics is that the concept of behaviour refers to an action on the part of the subject. On the contrary, gender, age or an IQ test result are considered personal attributes but not behavioural data. Roughly speaking, they tend to describe what the person *is* rather than what he or she *does*.[71] Behaviour is something for which a person could be held responsible. And the fact that the provision also refers to the "gravity" of the behaviour confirms this implicit view. It is as if the social scoring system was designed to sanction individuals for their *bad* behaviours. Given that the Chinese scoring system inspired the drafters of the AI Act, this is not surprising since the ambition of the Chinese system is indeed to make the behaviour of Chinese citizens more virtuous. For example, in the social credit system project proposed by the city of Changsha, actions such as volunteering with the Communist Youth League, donating to the Chinese Red Cross or giving blood increase your score; while actions such as not paying your employees' wages, being convicted of a crime or committing traffic violations make you lose points.[72]

---

[67] Vincent Caron, 'Le préjudice de perte de solvabilité découlant d'une inscription négative au dossier de crédit du consommateur' (2021) 50 Revue générale de droit 445. Our translation.
[68] Rona-Tas (n 66).
[69] Lauer (n 18) 172.
[70] As already said, this use of a score could also violate the GDPR, in particular the purpose limitation principle. We do not analyse this issue in more detail here, but it can be pointed out that there are exceptions to this principle (notably those set out in Article 6(4) of the GDPR), while Article 5(1)(c) of the AI Act does not suffer any exceptions. In addition, it should be noted that that the AI Act requires high-risk AI systems to specify their "intended purpose" as well as the "reasonably foreseeable misuse" (*i.e.* the use of an AI system for another purpose than its intended purpose) (see definitions at Article 3 (12) and (13)). The main requirement in this respect for the provider is to minimise (through the risk management system and human oversight) the risk of a "reasonably foreseeable misuse". The use of a credit score to select new employees could possibly be considered as such a reasonably foreseeable misuse. However, such misuse is not prohibited *per se*, but the associated risks must be mitigated.
[71] This distinction is of course far from clear-cut as it is often a daunting task to distinguish what is behavioural and what is not.

In Western countries, the system which is the closest to this idea of attributing points in relation to the good or bad behaviour of citizens is probably points-based driving licences: each driver loses points every time he or she infringes a traffic rule, which can ultimately lead to the withdrawal of the driving licence[73]. However, if we think of AI-enabled scoring, the idea that the variables influencing the score should be proportionate to the consequences resulting from the score sounds incongruous. AI developers would probably argue that scoring is not about punishing or rewarding someone for their good or bad actions, but about looking for attributes that can statistically predict unknown traits in order to improve decision-making processes. As discussed above, recent years have seen the emergence of new forms of scoring that take into account alternative data such as the time spent in reading the terms and conditions or the network of telephone contacts.[74] Is this proportionate to the "gravity" of the "behaviour" of the scored subject? Is it really justified and proportionate to deny someone access to credit because they are related to "bad borrowers"?[75] The way out for a bank taking into account such variable would probably be, first, to invoke the statistical validity of its model – if these criteria are right, it's because the data is correct and effectively predicts the risk –, and second to argue that it is only one variable among others that feed the model.

In any case, just as the criterion relating to the multiplicity of contexts of application, this condition is broad and makes it possible to tackle scoring practices which, even if they would not immediately be associated with social scoring, would nonetheless appear disproportionate.

---

[72] Arsène (n 8) LXV.
[73] In some systems, drivers do not lose points but accumulate them each time they commit a traffic offence (See José I Castillo-Manzano and Mercedes Castro-Nuño, 'Driving Licenses Based on Points Systems: Efficient Road Safety Strategy or Latest Fashion in Global Transport Policy? A Worldwide Meta-Analysis' (2012) 21 Transport Policy 191). It is worth noting that the behaviours penalised are prohibited by traffic regulations, whereas the Chinese system also rewards actions that are not compulsory but only considered virtuous.
[74] María Óskarsdóttir and others, 'The Value of Big Data for Credit Scoring: Enhancing Financial Inclusion Using Mobile Phone Data and Social Network Analytics' (2019) 74 Applied Soft Computing 26; Robinson + Yu, 'Knowing the Score: New Data, Underwriting, and Marketing in the Consumer Credit Marketplace. A Guide for Financial Inclusion Stakeholders' (2014) <https://www.upturn.org/static/files/Knowing_the_Score_Oct_2014_v1_1.pdf>.
[75] In a way, this requirement of justification and proportionality is linked to the "relatedness of context" criterion examined in 2.3.1. since taking into account data that are not related to the purpose of the evaluation can be considered unjustified and disproportionate.

## 5. Conclusion: a new tool to protect the scored subject?

Will European citizens be scored differently with the entry into force of the AI Act? The intuition that triggered the writing of this article is that the ban on social scoring is perceived mainly as a rule with little effect, adopted solely to avoid falling into a kind of authoritarian drift and which will not affect existing scoring practices. It's as if we had two types of scoring: on the one hand, scoring practices which already exist in Europe and certainly raise fundamental rights concerns, but which should nevertheless be promoted as long as they remain in line with the "human-centric and trustworthy approach to AI" that has been promoted by the EU;[76] on the other, a potential totalitarian scoring system that exerts harsh control over citizens by monitoring their most trivial behaviour and deciding every aspect of their lives. The analysis carried out in this article shows, however, that things are more complex, as the boundary between social scoring as defined in the AI Act and other lawful forms of scoring is far from sharp.

As has been demonstrated, the "social" nature of a social score within the meaning of Article 5(1)(c) is context dependant: it may become social because of how it is used. A social score can of course be produced by an AI system that is itself designed as a "social scoring system", i.e. a system which would be conceived from the outset to assess individuals in the light of various data and for different purposes. But the ecosystem of data and scores is such that data is collected, sold and transformed into scores, and the scores themselves are sold to third parties. In this sense, a score – for example, a credit score – can individually become a social score because it has been used in a way that qualifies it as "social" under the AI Act. In this regard, it is of great importance that Article 5(1)(c) does not ban social scoring *systems*, but only social scores. This finding echoes the conclusions of the CJEU in the *SCHUFA Holding AG Case* in which the Court stated that the score sold by SCHUFA was itself an automated decision because it played a decisive role in refusing to grant credit to the applicant. Even if the SCHUFA scoring system is not an automated decision-making system (since the scores do not necessarily involve a decision), the SCHUFA score was finally qualified as an automated decision (because of the influence of the score on the final refusal of credit *in this case*). Similarly, even if the SCHUFA scoring system (or any other consumer credit agency) is not a *social* scoring system, a SCHUFA score could qualify as a social score if it were used by a third-party in a different context than that of credit.

In addition, the ban on social scoring may also concern scores that are used for a specific purpose and in a limited context. Indeed, although Recital 31 states that lawful evaluation of individuals carried out for a specific purpose are not affected by this prohibition, the wording of Article 5(1)(c)(ii) specifies that a score shall be prohibited if the treatment resulting from this score is disproportionate or unjustified with regard to the social behaviour or its gravity. Nothing here prevents a score that is used for a specific purpose, but which appears disproportionate of unjustified from constituting a social score.

In light of this, the ban on social scoring appears to be a complementary tool to the ban on automated decisions introduced by Article 22 of the GDPR. Some practices that are not prohibited under Article 22 of the GDPR are now banned under Article 5(1)(c) (and, conversely, other practices that were already prohibited by Article 22 of the GDPR will remain so).[77] In some respects, Article 22 of the GDPR is broader in scope: it covers automated decisions (based on personal data), whereas Article 5(1)(c) covers AI-based scoring. Many scores are indeed not based on AI, especially as the AI Act's definition of AI systems requires them to operate with varying levels of autonomy and to show adaptiveness after deployment, which tends to narrow the scope of application of the AI Act.[78] In addition, Article 22 prohibits any automated scoring that leads to decisions producing legal or significant effects, whereas Article 5(1)(c) of the AI Act only prohibits AI-based scoring to the extent that the resulting detrimental treatment occurs in a different context or is disproportionate. However, the AI Act's ban on social scoring appears broader in one key aspect: unlike the GDPR's prohibition on automated decisions, it allows for no exceptions. While Article 22(2) of the GDPR permits automated decisions under certain conditions – when necessary for entering into a contract, when authorised by Member State law, or when based on the explicit consent of the data subject –, the ban on social scoring is absolute.

---

[76] For a critique of this approach: Derave, Genicot and Hetmanska (n 21).

From a fundamental rights perspective, the crucial point is therefore that the ban on social scoring confer a strict right on individuals to be protected against practices that may infringe their rights and be potentially harmful. In contrast, the safeguards put in place for high-risk AI systems are more vague and imprecise, as it is up to the providers and deployers of these systems to mitigate the risks they have identified themselves. The ban on social scoring could then be used to challenge the legality of common AI-enabled scoring techniques. When faced with an AI-enabled score, the question will arise as to the context in which the data was collected and the score produced, as well as the relationship of proportionality and necessity between the behaviour of the scored subject and the possible consequences of the score. It now remains to be seen whether the courts will be willing to embrace such a proactive interpretation of the ban on social scoring. The future will tell.

---

[77] In this regard, Article 2(7) of the AI Act states that the AI Act shall not affect EU data protection law (including the GDPR). However, this obviously does not mean that it cannot prohibit practices which are not already prohibited under EU data protection law (otherwise there would no point in banning such practices). The European Commission confirms this in its Guidelines on prohibited AI practices: "[The AI Act] provides additional protection by addressing specific harmful AI practices which may not be prohibited by other [EU] laws. (…) At the same time, the AI Act does not affect prohibitions that apply where an AI practice falls within other Union law. Thus, even where an AI system is not prohibited by the AI Act, its use could still be prohibited or unlawful based on other primary or secondary Union law (Guidelines on prohibited AI practices (n 15) 14, para 42-43).

[78] Article 3(1) of the AI Act. See also European Commission, 'Annex to the Communication to the Commission Approval of the content of the draft Communication from the Commission - Commission Guidelines on the definition of an artificial intelligence system established by Regulation (EU) 2024/1689 (AI Act)', C(2025) 924 final.